# Two-Dimensional Itinerant Ising Ferromagnetism in Atomically thin Fe$_3$GeTe$_2$

Zaiyao Fei[1†], Bevin Huang[1†], Paul Malinowski[1], Wenbo Wang[2], Tiancheng Song[1], Joshua Sanchez[1], Wang Yao[3], Di Xiao[4], Xiaoyang Zhu[5], Andrew May[6], Weida Wu[2], David Cobden[1], Jiun-Haw Chu[1]*, Xiaodong Xu[1,7]*

[1]Department of Physics, University of Washington, Seattle, Washington 98195, USA.
[2]Department of Physics and Astronomy, Rutgers University, Piscataway, New Jersey 08854, USA
[3]Department of Physics and Center of Theoretical and Computational Physics, University of Hong Kong, Hong Kong, China.
[4]Department of Physics, Carnegie Mellon University, Pittsburgh, Pennsylvania 15213, USA.
[5]Department of Chemistry, Columbia University, New York, New York 10027, USA.
[6]Materials Science and Technology Division, Oak Ridge National Laboratory, Oak Ridge, Tennessee 37831, USA.
[7]Department of Materials Science and Engineering, University of Washington, Seattle, Washington 98195, USA.

[†]These authors contributed equally to this work.
*Correspondence to: xuxd@uw.edu, jhchu@uw.edu

**Abstract:** Recent discoveries of intrinsic two-dimensional (2D) ferromagnetism in insulating/semiconducting van der Waals (vdW) crystals open up new possibilities for studying fundamental 2D magnetism and devices employing localized spins[1–4]. However, a vdW material that exhibits 2D itinerant magnetism remains elusive. In fact, the synthesis of such single-crystal ferromagnetic metals with strong perpendicular anisotropy at the atomically thin limit has been a long-standing challenge. Here, we demonstrate that monolayer Fe$_3$GeTe$_2$ is a robust 2D itinerant ferromagnet with strong out-of-plane anisotropy. Layer-dependent studies reveal a crossover from 3D to 2D Ising ferromagnetism for thicknesses less than 4 nm (five layers), accompanying a fast drop of the Curie temperature from 207 K down to 130 K in the monolayer. For Fe$_3$GeTe$_2$ flakes thicker than ~15 nm, a peculiar magnetic behavior emerges within an intermediate temperature range, which we show is due to the formation of labyrinthine domain patterns. Our work introduces a novel atomically thin ferromagnetic metal that could be useful for the study of controllable 2D itinerant Ising ferromagnetism and for engineering spintronic vdW heterostructures[5].

**Main Text**

Bulk crystalline iron germanium telluride ($Fe_3GeTe_2$, or FGT, see Fig. 1a) is an itinerant van der Waals (vdW) ferromagnet with a reasonably high Curie temperature ($T_C$) of about 220-230 K[6–10]. In addition to a large out-of-plane anisotropy[8], evidence for strong electron correlation effects[11] and Kondo lattice physics[12] has been seen recently. Encouragingly, first-principles calculations indicate that the large perpendicular magnetic anisotropy will persist down to a monolayer[13], supporting the possibility of sustaining long-range magnetic order in the two-dimensional limit[14]. In fact, anomalous Hall effect (AHE) measurements on epitaxially grown FGT thin fims[9] have revealed only a slight reduction of $T_C$ from that of the bulk, further supporting the likelihood of anisotropy-stabilized long-range ferromagnetic order in ultrathin FGT flakes. The combination of these desirable properties makes FGT a promising candidate for exploring, for the first time, itinerant ferromagnetism in a truly 2D form.

In this work, FGT flakes down to the monolayer limit were prepared by mechanical exfoliation from bulk crystals. The magnetic order was probed with both magneto-transport and Kerr rotation measurements. For the former, we cleaved FGT bulk crystals onto 285 nm $SiO_2$/Si substrates and patterned the resulting exfoliated thin flakes into Hall bars using electron beam lithography (see Methods). Atomic force microscopy (AFM) measurements on a stepped FGT flake show the layer step height to be ~0.8 nm (Fig. 1b), consistent with X-ray diffraction measurements[8]. The thinnest flake we could obtain by exfoliation onto $SiO_2$/Si was 4.8 nm, i.e., 6 layers. For even thinner samples, we cleaved FGT bulk crystals onto a gold film evaporated on top of $SiO_2$/Si. As demonstrated with other materials including $MoS_2$[15–17] and $Bi_2Te_3$[15], the gold substrate gives an increased yield of atomically thin flakes. However, the magnetic order in such samples can only be characterized by optical methods.

We first examine the magnetic properties of FGT thin flakes through magneto-transport measurements. Figure 1c shows the temperature-dependent longitudinal resistance, $R_{xx}$(T), of a representative 12-nm FGT Hall bar device (optical micrograph in upper-left inset). At ~210 K, the kinks in both $R_{xx}$ and its first temperature derivative (lower-right inset) are characteristic of the established phase transition from paramagnetism to ferromagnetism. The Hall resistance, $R_{xy}$, shown in Fig. 1d, shows rectangular hysteresis loops with near-vertical jumps at low temperatures as a function of external magnetic field, $\mu_oH$, applied perpendicular to the sample plane, with the coercive field reaching 1 T at 2 K. This indicates the domination of the anomalous Hall effect with a single magnetic domain over the entire flake. Together with the large remanent $R_{xy}$ at zero field, these are hallmarks of ferromagnetism with strong out-of-plane anisotropy. The remanent $R_{xy}$ and coercive field vanishes at around 210 K, consistent with the kinks in Fig. 1c (see S1 for additional magneto-transport measurements). These transport measurements show that the itinerant ferromagnetism exhibited in bulk crystals[6–8] and in MBE grown samples[9] persists in thin exfoliated FGT flakes.

To study truly 2D itinerant ferromagnetism, we investigated FGT flakes down to the monolayer limit on the gold substrates (see Fig. S2 for optical and AFM images). We employed both polar reflective magnetic circular dichroism (RMCD) and magneto-optical Kerr effect (MOKE) microscopy to examine the out-of-plane magnetic order (see Methods). The magnetic field was applied perpendicular to the sample and the excitation laser was a 1-μW 633 nm HeNe with a spot size of 3 μm. Figure 2a shows the RMCD signal of a monolayer (false-color optical micrograph in inset) as $\mu_oH$ was swept between ±0.2 T at 78 K. Remarkably, ferromagnetism persists, as evidenced by the prominent hysteresis seen here. A spatial map of the RMCD intensity at 0.2 T (Fig. 2b) also reveals the magnetization to be uniform across the entire monolayer flake.

Figure 2c shows RMCD measurements as a function of magnetic field for several fixed temperatures. As the temperature increases, the hysteresis loop shrinks, disappearing at about 130 K. The same behavior was observed for all four monolayers measured and implies that $T_C$, the temperature at which the ferromagnetic phase transitions to the paramagnetic phase, is about 130 K for monolayers. The $T_C$ can also be extracted from a temperature dependence of the remanent RMCD signal (blue circles in Fig. 3). The onset of magnetization in the monolayer flake clearly occurs at around 130 K, a substantial decrease from the bulk value but nevertheless significantly higher than that the transition temperatures seen in exfoliated ferromagnetic monolayers of $CrI_3$ (45 K)[1] and bilayers of $Cr_2Ge_2Te_6$ (30 K at 0.075 T)[2].

On FGT flakes of up to five-layers, the dependence of $T_C$ on number of layers is consistent with a dimensional crossover taking place between the atomically thin limit and the bulk crystal. In Fig. 3, we compare the temperature dependence of the remanent RMCD signal for monolayer, bilayer (dark green), trilayer (light green), four-layer (orange), and five-layer (red) flakes. For four and five layers, the $T_C$ is only slightly below the 3D value of around 220 K. In contrast, for fewer layers, the $T_C$ sharply declines, dropping to 180 K in the bilayer and 130 K in the monolayer. This behavior resembles the thickness-dependent $T_C$ associated with dimensional crossover observed in ultrathin magnetic metal systems[18,19].

Such a crossover can be confirmed by finding the critical exponent, $\beta$, associated with the temperature dependence of the magnetization[18–20]. For each sample, we fitted the remanent RMCD signal to the critical power-law form $(1 - T/T_C)^\beta$ using $\beta$ and $T_C$ as two simultaneous fitting parameters. The values of $\beta$ obtained are plotted versus thickness in the inset to Fig. 3, including results from thicker flakes up to 18 nm (see Fig. S3). (Note that in these ultra-thin flakes there is usually only a single magnetic domain.) For the thicker flakes, $\beta$ is in the range $0.25 - 0.27$, compared with the values of 0.327 (ref. 21) and 0.25 (ref. 7) reported for bulk crystalline FGT and the 3D Ising value of 0.33. For thinner flakes, $\beta$ falls to $0.2 \pm 0.02$ in five- to bilayer flakes, and to $0.14 \pm 0.02$ in the monolayer, consistent with $\beta = 0.125$ for the 2D Ising model.

In the few-layer samples, the deviation from the 3D Ising exponent could be due to finite sample size[19,22], limitations on temperature control (about ±1 K in our experiment), and the fact that the fits are performed over a relatively wide temperature range exceeding the width of the critical region. A crossover from 3D to 2D critical behavior should also occur as a function of reduced temperature on a scale set by the sample thickness, which makes fitting to a single power law invalid. Nevertheless, the substantial reduction of $\beta$ to the 2D Ising value and the accompanying large drop in $T_C$, combined with the large out-of-plane anisotropy, altogether indicate that FGT monolayers are truly 2D itinerant Ising ferromagnets.

The atomically-thin flakes discussed so far exhibited rectangular hysteresis loops, indicative of single-domain, highly anisotropic out-of-plane ferromagnetic ordering. However, for thicker FGT flakes, the hysteresis behavior changes notably. Figures 4a&b compare magnetic field sweeps at the same fixed temperatures for a 3.2 nm (four-layer) flake and a 48 nm flake. (See S4 for measurements on other thicknesses). The 3.2-nm flake exhibits a rectangular hysteresis loop at all temperatures below $T_C = 204$ K, while above $T_C$ the hysteresis and remanent RMCD signal vanish. In contrast, the hysteresis loop of the 48-nm FGT flake is only rectangular below 185 K, while paramagnetism sets in, with vanishing hysteresis, above $T_C$ ~210 K. However, within an intermediate temperature range (185 – 210 K), the hysteresis loops differ from that of either state. When the field is swept upwards starting at -1 T, the RMCD signal suddenly jumps from the saturation value to an intermediate level at a certain magnetic field $B_{in}$ (see the grey arrow in Fig. 4b as an example). The RMCD then changes roughly linearly, passing through zero at zero field, and finally saturates at high positive field. The time-reversed process does the opposite, as expected. As temperature increases, the magnitude of $B_{in}$ increases until it merges with the saturation field at $T_C$. AHE measurements show exactly the same behavior (Fig. S5).

A possible explanation for this behavior is the formation of domain structures that are unresolvable through our optical measurements. Bulk crystal FGT exhibits stripy and bubble-like domains[10,23,24]. On the other hand, Co/Pt multilayer films with low disorder[25] show similar hysteresis loops to those in Fig. 4b, known to be due to the formation of labyrinthine magnetic domains. Theoretical calculations for these magnetic films have reproduced the experimental results remarkably well, accurately predicting the domain size for a given film thickness and the critical thickness at which the domains appear[26,27]. The striking similarities to our observations suggests that labyrinthine domains are involved.

To confirm this, we performed magnetic force microscopy (MFM) on a ~340-nm thick flake at several fixed magnetic fields (Fig. S6). Figure 4c shows the MFM image at zero magnetic field and 170 K, clearly revealing labyrinthine domains, which appear suddenly when the field is reduced through $|B_{in}|$ (Fig. S6d). A line cut along the black dashed line in Fig. 4c shows domain widths of ~200 nm (Fig. 4d). We deduce that the RMCD signal, with a beam spot size of ~3 μm, averages over the out-of-plane magnetization contributions from many domains. The behavior in Fig. 4b is therefore due to the appearance of domains at $|B_{in}|$ followed by their gradual evolution, giving a steady rate of net magnetization change, until the saturation field is reached, and the sample returns to a uniform magnetization state. The similarity with the Co/Pt multilayer films further supports the conclusion that FGT is a low-disorder metal with out-of-plane ferromagnetism[25].

Figure 4e is a thickness-temperature phase diagram for the three distinct magnetic states–paramagnetic (PM), single-domain ferromagnetic (FM1), and labyrinthine-domain ferromagnet (FM2), compiled from all samples of different thicknesses. Below ~15 nm, there is only a ferromagnetic to paramagnetic transition at a single temperature $T_C$. Above the same thickness, labyrinthine domains occur in a temperature range from $T_{c1}$ (lower) to $T_{c2}$ (upper). According to modeling[27], this temperature-driven magnetic domain formation implies a decrease of the ratio of the strength of out-of-plane anisotropy to that of exchange interactions as the temperature increases.

In summary, the van der Waals layered magnet FGT provides a model system for layered itinerant ferromagnetism, showing various layer-number dependent magnetic phenomena. Monolayers of FGT exhibit true 2D Ising ferromagnetism and could potentially be electrically gated to allow tuning of the magnetic properties. $Fe_3GeTe_2$ could also be employed for ferromagnetic contacts for injecting spins to other 2D materials, such as topological insulators, 2D valley semiconductors, and 2D superconductors, for the creation and investigation of emerging physical phenomena and heterostructure spintronics.

**Methods**

**Crystal Growth and Characterization:** Bulk single-crystals of $Fe_3GeTe_2$ were grown by the chemical vapor transport method using iodine as a transport agent[7]. Elemental powders of high purity Fe (99.998%), Ge (99.999%), and Te (99.999%) were weighed out in stoichiometric ratios of 3:1:2 and cold pressed into a single pellet. This pellet was then placed along with 2 mg/cm$^3$ of solid $I_2$ into a quartz tube which was evacuated of air and brought to 15 millitorr argon atmosphere before sealing. The sealed tube was then set in a temperature gradient of 750/650° C for one week, with the starting materials placed at the hot end. Single crystals precipitated on the cold end of the tube. We performed energy dispersive X-ray spectroscopy (EDS) on crystal in order to determine the molar ratio of the grown crystals, which averaged over multiple samples to be 3:1.1:1.9.

**Sample Preparation:** For magneto-transport, RMCD, and MOKE measurements, $Fe_3GeTe_2$ crystals were cleaved inside an inert gas glovebox (oxygen and water vaper levels below 0.5 ppm) and deposited directly onto 285 nm $SiO_2$/Si substrates. Thin flakes (5-300 nm thick) were identified based on their optical contrasts and their thicknesses confirmed through atomic force microscopy (AFM). Devices were formed from these flakes by patterning and evaporating V/Au contacts in a Hall bar geometry through standard electron-beam lithography procedures and electron-beam evaporation.

Since the yield of atomically thin flakes (thinner than 5 nm) is significantly low on $SiO_2$ substrates, we first evaporated 5/12 nm V/Au films on top of the 285 nm $SiO_2$/Si substrates and exfoliated $Fe_3GeTe_2$ on these films. Atomically thin flakes down to a monolayer were identified and confirmed through AFM, which were then studied through RMCD/MOKE microscopy.

**Electrical Measurements:** All magneto-transport measurements were carried out in a Quantum Design PPMS (DynaCool) cryostat with temperatures down to 2 K and magnetic fields of up to 9 T. A constant 1 µA AC current at 13 Hz was applied for all Hall measurements.

**MOKE and RMCD Measurements:** A material that exhibits a non-zero magnetic moment may also display magnetic circular birefringence (MCB) and magnetic circular dichroism (MCD), with the former leading to an overall phase difference between right-circularly polarized (RCP) light and left-circularly polarized (LCP) light and the latter resulting in an amplitude difference between RCP and LCP. When linearly polarized light, an equal superposition of LCP and RCP light, reflects off this magnetic material, the linear polarization rotates through an angle known as the Kerr rotation from MCB and becomes elliptically polarized from MCD. These are known as the magneto-optical Kerr effect (MOKE) and reflective magnetic circular dichroism (RMCD), respectively.

MOKE and RMCD measurements were performed in a closed-cycle helium cryostat with a temperature range from 15-300 K and an out-of-plane magnetic field of up to 7 T. A power-stabilized 633 nm HeNe laser of about 1 µW was focused to a 3 µm beam spot on the sample at normal incidence. The optical setup follows closely to previous MOKE measurements on $CrI_3$[1]. RMCD measurements were taken with the analyzing polarizer removed and with the lock-in amplifier set to the PEM frequency.

**MFM:** The MFM experiments were carried out in a homemade cryogenic atomic force microscope (AFM) using commercial piezoresistive cantilevers (spring constant $k \approx 3$ N/m, resonant frequency $f_0 \approx 42$ kHz). The homemade AFM is interfaced with a Nanonis SPM Controller using a phase-lock loop (SPECS). MFM tips were prepared by depositing a 150-nm Co film onto bare tips using e-beam evaporation. MFM images were taken in a constant height mode with the scanning plane ~160 nm above the sample surface. The MFM signal, the change of the cantilever resonant frequency, $\delta f$, is proportional to the out-of-plane stray field gradient.

**Acknowledgments:** The authors thank Marcel den Nijs for the helpful discussion. The work at UW is mainly supported by NSF MRSEC 1719797. BH and DX are supported by Basic Energy Sciences, Materials Sciences and Engineering Division (DE-SC0012509). WY is supported by the Croucher Foundation (Croucher Innovation Award) and the HKU ORA. Synthesis efforts at ORNL (AFM) were supported by the US Department of Energy, Office of Science, Basic Energy Sciences, Materials Sciences and Engineering Division. XX and XYZ acknowledge NSF MRSECs at UW (DMR-1719797) and Columbia (DMR-1420634) for supporting the exchange visit. XX and JHC acknowledges the support from the State of Washington funded Clean Energy Institute and from the Boeing Distinguished Professorship in Physics. Work at Rutgers (WWa and WWu) was supported by the Office of Basic Energy Sciences, Division of Materials Sciences and Engineering, U.S. Department of Energy under Award number DE-SC0018153.

**Author Contributions:** XX, JHC, and AM conceived the experiment. PM and AM synthesized and characterized the bulk $Fe_3GeTe_2$ crystal, assisted by JS. XYZ designed the exfoliation on gold substrate approach. ZF and BH fabricated the samples, acquired the experimental data, assisted by TS, supervised by XX, JHC, and DHC. WWa and WWu performed and analyzed MFM results. DX and WY provided theoretical support. ZF, BH, XX, JHC, and DHC wrote the manuscript with input from all authors. All authors discussed the results.

**References:**

1. Huang, B. *et al.* Layer-dependent ferromagnetism in a van der Waals crystal down to the monolayer limit. *Nature* **546,** 270–273 (2017).

2. Gong, C. *et al.* Discovery of intrinsic ferromagnetism in two-dimensional van der Waals crystals. *Nature* **546,** 265–269 (2017).

3. Seyler, K. L. *et al.* Ligand-field helical luminescence in a 2D ferromagnetic insulator. *Nat. Phys.* (2017). doi:10.1038/s41567-017-0006-7

4. Bonilla, M. *et al.* Strong room-temperature ferromagnetism in $VSe_2$ monolayers on van der Waals substrates. *Nat. Nanotechnol.* (2018). doi:10.1038/s41565-018-0063-9

5. Zhong, D. *et al.* Van der Waals engineering of ferromagnetic semiconductor heterostructures for spin and valleytronics. *Sci. Adv.* **3,** e1603113 (2017).

6. Deiseroth, H.-J., Aleksandrov, K., Reiner, C., Kienle, L. & Kremer, R. K. $Fe_3GeTe_2$ and $Ni_3GeTe_2$ – Two New Layered Transition-Metal Compounds: Crystal Structures, HRTEM Investigations, and Magnetic and Electrical Properties. *Eur. J. Inorg. Chem.* **2006,** 1561–1567 (2006).

7. Chen, B. *et al.* Magnetic Properties of Layered Itinerant Electron Ferromagnet $Fe_3GeTe_2$. *J. Phys. Soc. Japan* **82,** 124711 (2013).

8. May, A. F., Calder, S., Cantoni, C., Cao, H. & McGuire, M. A. Magnetic structure and phase stability of the van der Waals bonded ferromagnet $Fe_{3-x}GeTe_2$. *Phys. Rev. B* **93,** 014411 (2016).

9. Liu, S. *et al.* Wafer-scale two-dimensional ferromagnetic $Fe_3GeTe_2$ thin films were grown by molecular beam epitaxy. *npj 2D Mater. Appl.* **1,** 30 (2017).

10. Yi, J. *et al.* Competing antiferromagnetism in a quasi-2D itinerant ferromagnet: $Fe_3GeTe_2$. *2D Mater.* **4,** 011005 (2016).

11. Zhu, J. X. *et al.* Electronic correlation and magnetism in the ferromagnetic metal $Fe_3GeTe_2$. *Phys. Rev. B* **93,** 144404 (2016).

12. Zhang, Y. *et al.* Emergence of Kondo lattice behavior in a van der Waals itinerant ferromagnet, $Fe_3GeTe_2$. *Sci. Adv.* **4,** eaao6791 (2018).

13. Zhuang, H. L., Kent, P. R. C. & Hennig, R. G. Strong anisotropy and magnetostriction in the two-dimensional Stoner ferromagnet $Fe_3GeTe_2$. *Phys. Rev. B* **93,** 134407 (2016).

14. Mermin, N. D. & Wagner, H. Absence of ferromagnetism or antiferromagnetism in one- or two-dimensional isotropic Heisenberg models. *Phys. Rev. Lett.* **17,** 1133–1136 (1966).

15. Magda, G. Z. *et al.* Exfoliation of large-area transition metal chalcogenide single layers. *Sci. Rep.* **5,** 14714 (2015).

16. Hsu, C. L. *et al.* Layer-by-layer graphene/TCNQ stacked films as conducting anodes for organic solar cells. *ACS Nano* **6,** 5031–5039 (2012).

17. Desai, S. B. *et al.* Gold-Mediated Exfoliation of Ultralarge Optoelectronically-Perfect

Monolayers. *Adv. Mater.* **28,** 4053–4058 (2016).

18. Li, Y. & Baberschke, K. Dimensional crossover in ultrathin Ni(111) films on W(110). *Phys. Rev. Lett.* **68,** 1208–1211 (1992).

19. Huang, F., Kief, M. T., Mankey, G. J. & Willis, R. F. Magnetism in the few-monolayers limit: A surface magneto-optic Kerr-effect study of the magnetic behavior of ultrathin films of Co, Ni, and Co-Ni alloys on Cu(100) and Cu(111). *Phys. Rev. B* **49,** 3962–3971 (1994).

20. Back, C. H. *et al.* Experimental confirmation of universality for a phase transition in two dimensions. *Nature* **378,** 597–600 (1995).

21. Liu, B. *et al.* Critical behavior of the van der Waals bonded high $T_C$ ferromagnet $Fe_3GeTe_2$. *Sci. Rep.* **7,** 6184 (2017).

22. Huang, F., Mankey, G. J., Kief, M. T. & Willis, R. F. Finite-size scaling behavior of ferromagnetic thin films. *J. Appl. Phys.* **73,** 6760–6762 (1993).

23. León-Brito, N., Bauer, E. D., Ronning, F., Thompson, J. D. & Movshovich, R. Magnetic microstructure and magnetic properties of uniaxial itinerant ferromagnet $Fe_3GeTe_2$. *J. Appl. Phys.* **120,** 083903 (2016).

24. Nguyen, G. D. *et al.* Visualization and manipulation of magnetic domains in the quasi-two-dimensional material $Fe_3GeTe_2$. *Phys. Rev. B* **97,** 014425 (2018).

25. Pierce, M. S. *et al.* Disorder-induced microscopic magnetic memory. *Phys. Rev. Lett.* **94,** 017202 (2005).

26. Deutsch, J. M. & Mai, T. Mechanism for nonequilibrium symmetry breaking and pattern formation in magnetic films. *Phys. Rev. E* **72,** 016115 (2005).

27. Jagla, E. A. Hysteresis loops of magnetic thin films with perpendicular anisotropy. *Phys. Rev. B* **72,** 094406 (2005).

Figures:

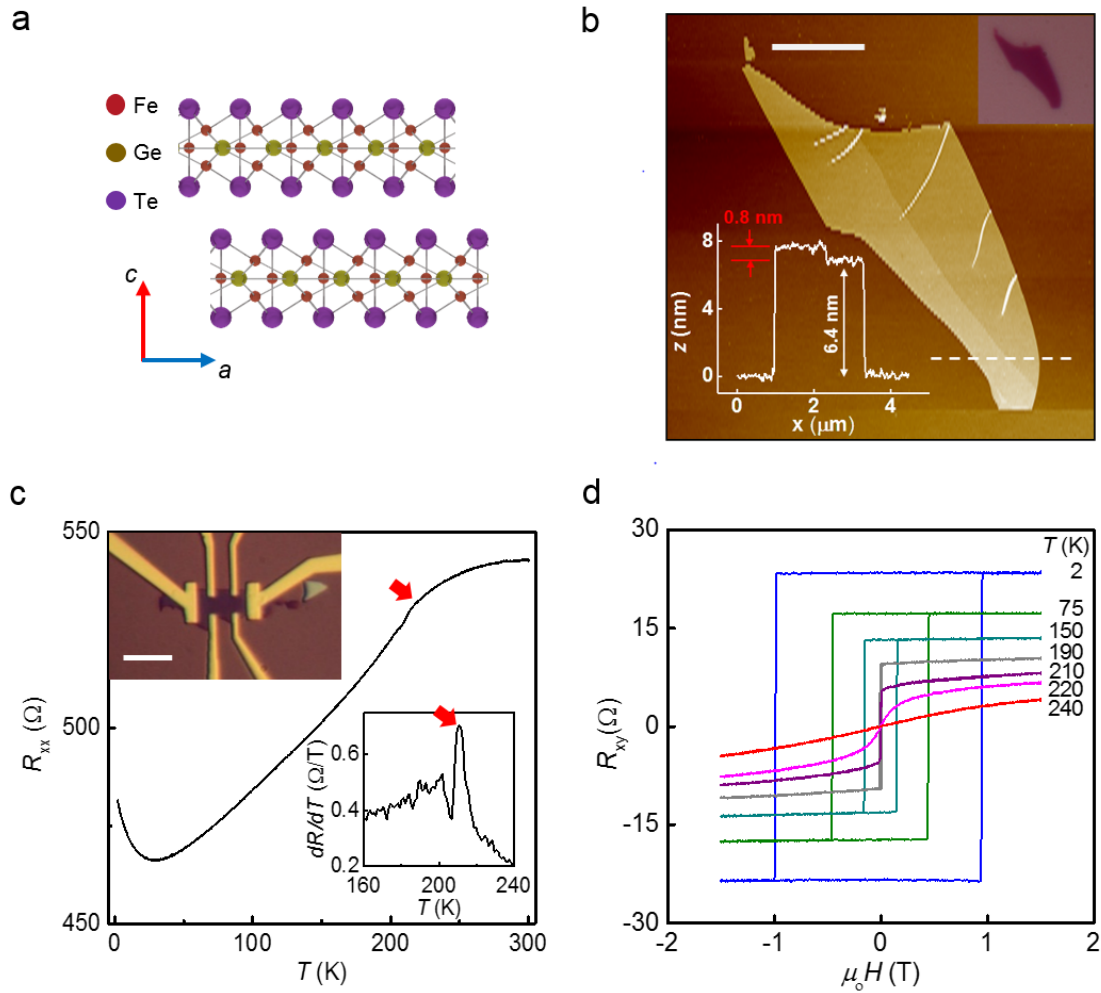

**Figure 1 | Structure and transport characterization of thin Fe₃GeTe₂ (FGT) flakes. a,** Side view of the atomic lattice of bilayer FGT. Each colored ball represents iron (red), germanium (yellow), and tellurium (purple). **b,** AFM (main) and optical (upper right inset) images of a representative thin FGT flake on 285 nm SiO$_2$. The scale bar is 3 μm. The lower-left inset shows a line cut along the white dashed line where a monolayer step of ~0.8 nm can be seen. **c,** Temperature dependence of the longitudinal resistance of a representative FGT device (12-nm thick). The upper-left inset shows an optical image of the Hall-bar device. The scale bar is 5 μm. The bottom-right inset shows the first derivative of the longitudinal resistance as a function of temperature. The red arrows show where the ferromagnetic-paramagnetic transition occurs. **d,** Temperature-dependent magnetic field (out-of-plane) sweeps of the Hall resistance measured on the 12-nm thick FGT device.

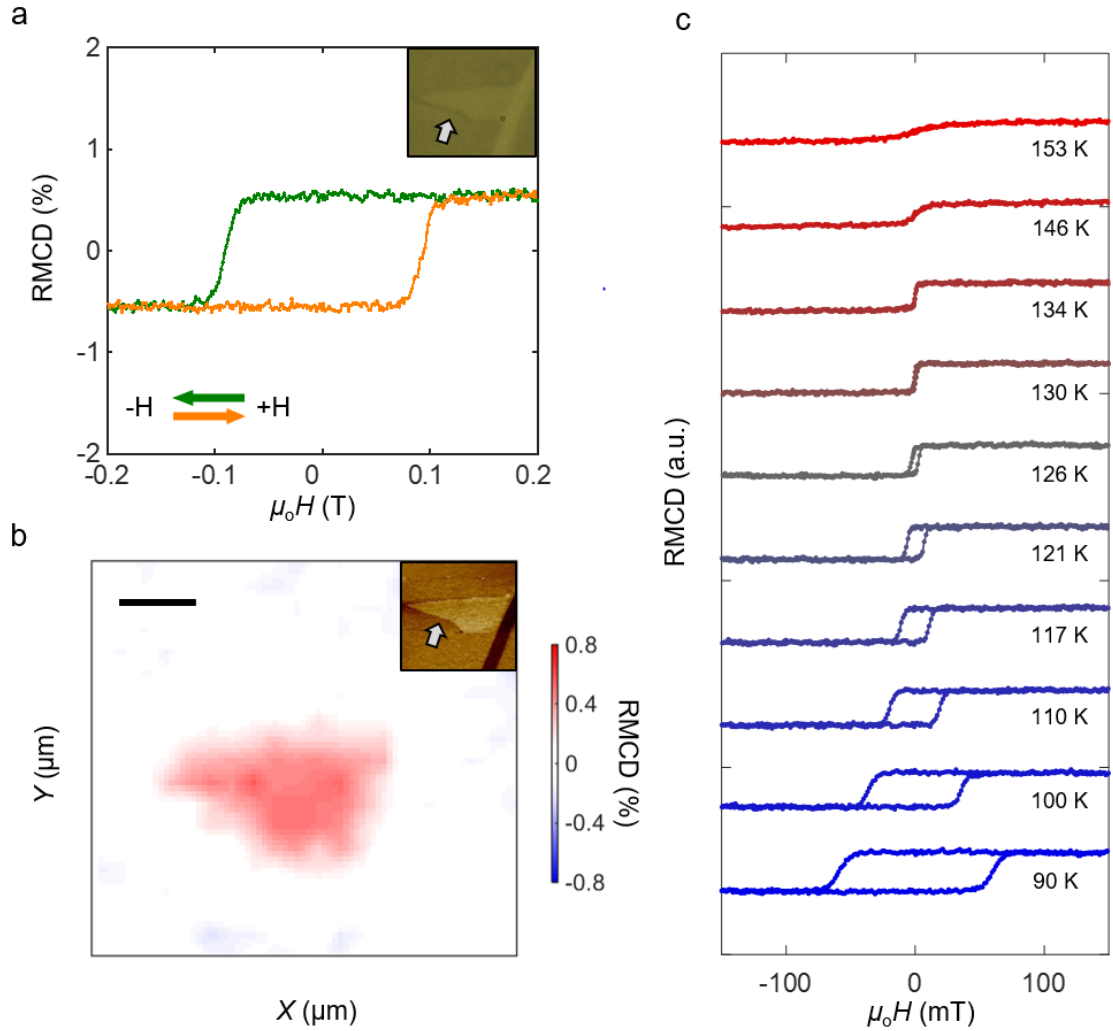

**Figure 2 | RMCD measurements of a monolayer FGT. a,** Polar RMCD signal for a monolayer FGT measured at 78 K. The inset is an optical image of the sample. **b,** RMCD map at $\mu_oH = 0.2$ T of the same sample at 78 K. The inset is an AFM image. The scale bar is 3 μm. The white arrows in **a** and **b** point to the monolayer flake. **c,** Magnetic field sweeps for the same sample at a range of temperatures passing through $T_C \sim 130$ K. The coercive field and remanent magnetization decreases as the temperature increases, until at 130 K both vanish.

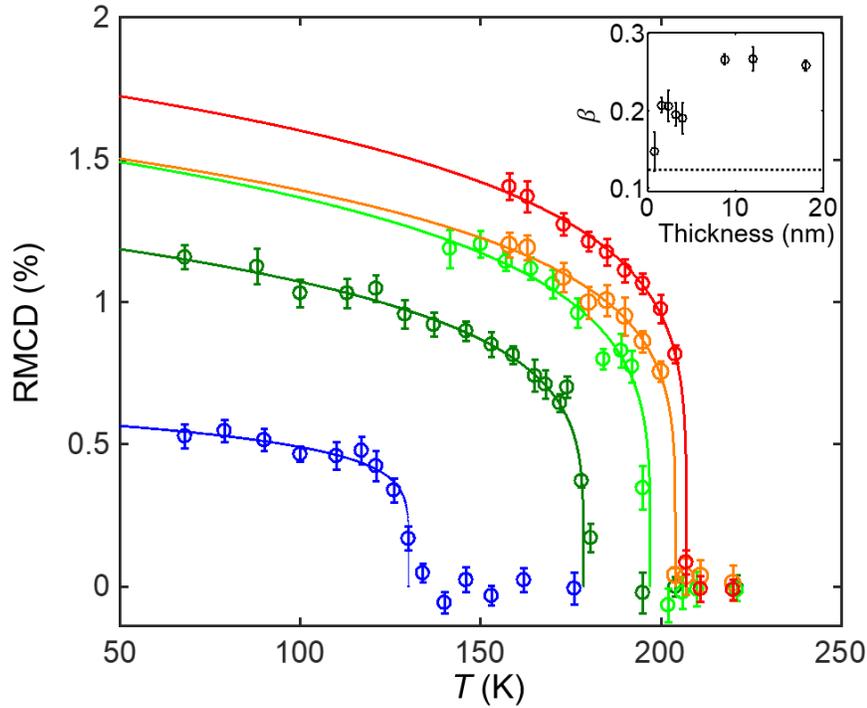

**Figure 3 | Criticality analysis for FGT flakes of different thicknesses.** The main panel shows the remanent RMCD signal as a function of temperature for a sequence of selected few-layer flakes (blue: monolayer; dark green: bilayer; pale green: trilayer; dark yellow: 4-layer; red: 5-layer). The error bars correspond to the standard deviation of the noise in the RMCD signal. The solid lines are least-squares criticality fits of the form: $\alpha(1 - T/T_C)^\beta$. Inset: derived values of the exponent $\beta$ plotted as a function of thickness, indicating a dimensionality crossover from 3D to 2D Ising ferromagnetism. The dotted line denotes $\beta = 0.125$, the critical exponent for the 2D Ising model. The error bars indicate the standard deviations in the least-squares fitting. Additional fits of $\beta$ for the 9-, 12- and 18-nm flakes are in Fig. S3.

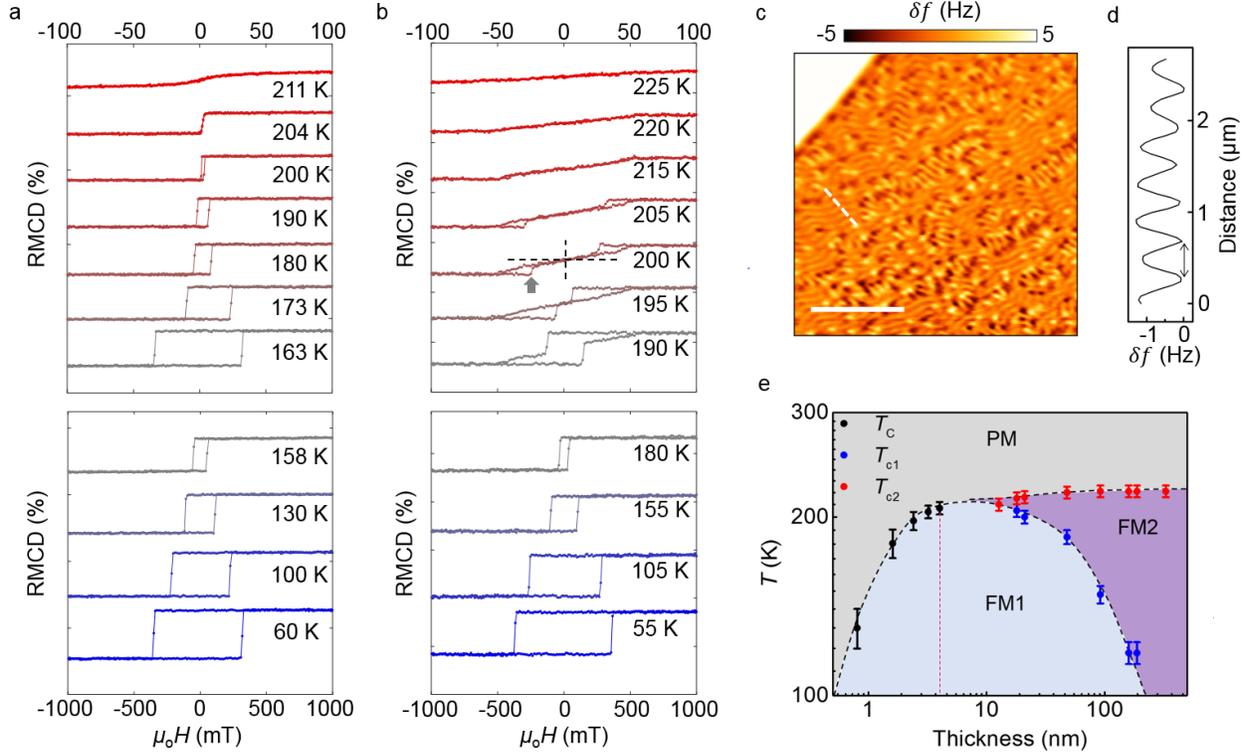

**Figure 4 | Intermediate magnetic states and thickness dependent phase diagram of FGT. a-b,** Comparison of RMCD sweeps for FGT of (a) 3.2-nm and (b) 48-nm thick. In the 200 K data in (b), the gray arrow indicates the sudden jump in RMCD at $-B_{in}$ as described in the main text, and the black dashed vertical and horizontal lines denote zero magnetic field and zero RMCD signal respectively. **c,** Magnetic force microscopy (MFM) image of a 340-nm thick FGT at 170 K, performed at $\mu_o H = 0$ T, showing labyrinthine domain structures. The scale bar is 5 μm. **d,** Line cut along the white dashed line in **c**, from which the domain size is identified to be ~200 nm. The detuning frequency from the AFM tip frequency, $\delta f$, is proportional to the out-of-plane stray field gradient. **e,** Compiled thickness-temperature phase diagram. Here, PM denotes the region in which the flake is paramagnetic, FM1 when the flake is ferromagnetic with a single-domain, and FM2 when the flake exhibits labyrinthine domains. The error bars are the uncertainties in determining the transition temperatures, $T_C$, $T_{c1}$, and $T_{c2}$, based on the temperature-dependent RMCD or AHE measurements of each flake thickness. The red dashed line denotes the critical thickness at which the dimensional crossover occurs. See text for details of the phase diagram.